\documentstyle[aps,epsf]{revtex}

\tighten
\newcommand{\beq}{\begin{equation}}
\newcommand{\eeq}{\end{equation}}
\newcommand{\ba}{\begin{eqnarray}}
\newcommand{\ea}{\end{eqnarray}}

\newcommand{\dz}{\int \frac{d^{4}z}{(2\pi)^4}}

\newcommand{\bpt}{\bm p_T^{}}
\newcommand{\bkt}{\bm k_T^{}}
\newcommand{\psibar}{\overline{\psi}}
\newcommand{\la}{\langle}
\newcommand{\ra}{\rangle}
\newcommand{\amp}[1]{\la #1 \ra}

\newcommand{\slsh}[1]{\mbox{$\not\! #1$}}
\newcommand{\bm}[1]{\bbox{#1}}

\begin{document}
 
\draft
\title{
\begin{flushright}
\begin{minipage}{4 cm}
\small
hep-ph/9902255\\
RIKEN-BNL preprint
\end{minipage}
\end{flushright}
Investigating the origins of transverse spin asymmetries at RHIC}

\author{Dani\"el Boer}
\address{\mbox{}\\
RIKEN-BNL Research Center\\
Brookhaven National Laboratory, Upton, New York 11973}

\maketitle
\begin{center}\today \end{center}

\begin{abstract}
We discuss possible origins of transverse spin asymmetries in hadron-hadron
collisions and propose an explanation in terms of a chiral-odd T-odd 
distribution function with intrinsic transverse momentum dependence, which 
would signal a correlation between the transverse spin and the 
transverse momentum of quarks inside an unpolarized hadron.
We will argue that despite its conceptual problems, it can account for 
single spin asymmetries, for example in 
$p \, p^{\uparrow} \rightarrow \pi \, X$, and at the same time for the large 
$\cos 2\phi$ asymmetry in the unpolarized Drell-Yan cross section, 
which still lacks understanding. We use the latter asymmetry to arrive at
a crude model for this function and show explicitly how it relates 
unpolarized and polarized observables in the Drell-Yan process, as 
could be measured with the proton-proton collisions at RHIC. 
Moreover, it would provide an alternative method of 
accessing the transversity distribution function $h_1$. 
For future 
reference we also list the complete set of azimuthal asymmetries in the 
unpolarized and polarized Drell-Yan 
process at leading order involving T-odd distribution functions with intrinsic
transverse momentum dependence.   
\end{abstract}

\pacs{13.88.+e; 13.85.Qk}  

\section{Introduction}
Large single transverse spin asymmetries have been observed experimentally
in the process $p \, p^{\uparrow} \rightarrow \pi \, X$ \cite{Adams} and 
many theoretical
studies have been devoted to explain the possible origin(s) of such 
asymmetries. However, one experiment cannot reveal the origin(s) 
conclusively and one needs comparison to other experiments. 
In this article we will use additional experimental results to propose an
explanation in terms of a chiral-odd T-odd distribution function with 
intrinsic transverse momentum dependence and we contrast it to the 
more standard theoretical proposals (cf.\ \cite{Boros}). In addition, it 
can account for the large 
$\cos 2\phi$ asymmetry in the unpolarized Drell-Yan cross section 
\cite{NA10,Conway}, which still lacks understanding. 

Unlike the chiral-even T-odd distribution function with 
intrinsic transverse momentum dependence as investigated by
\cite{s90,Anselmino}, which 
depends on the polarization of the parent hadron, the
chiral-odd function signals a correlation between the transverse spin and the
transverse momentum of quarks inside an {\em unpolarized\/} hadron. 
But one can
use the polarization of one hadron to become sensitive to the polarization
of quarks in another, unpolarized hadron. In this way it could provide a new 
way of 
measuring the transversity distribution function $h_1$. We propose two 
measurements that could be done at RHIC 
using polarized proton-proton collisions to study  
such a mechanism and to try to obtain information on $h_1$. 
 
Apart from discussing the advantages of this proposal, we will discuss the 
theoretical difficulties connected to such a function. The function is
actually the distribution function analog of the fragmentation function
associated with the Collins effect \cite{Collins-93b}. Unlike the fragmentation
function the distribution function is expected to be zero due to time reversal
symmetry, unless one assumes some nonstandard mechanism to generate such a
function, like for instance  
factorization breaking, which implies non-universality, or 
effects due to the finite size of a hadron, in which case one has the 
problem of systematically taking into account such effects in hard 
scattering factorization.

The article is organized as follows. In Section II we discuss possible origins
for transverse spin asymmetries. In Section III we elaborate on 
transverse momentum dependent distribution functions. In Section IV and also
in the appendix, we give 
results for the leading order Drell-Yan cross section in 
terms of the T-odd distribution 
functions, for completeness taking into account contributions 
from Z-bosons. In Section V we discuss how one particular 
function can not only explain (in principle) the single
spin asymmetries in the process $p \, p^{\uparrow} \rightarrow \pi \, X$, but
also explain the azimuthal $\cos 2 \phi$ dependence of the unpolarized
Drell-Yan cross section data \cite{NA10,Conway}. In Section VI we propose
the measurements that could be performed at RHIC and that might uncover such
an underlying mechanism.
In Section VII we discuss the conceptual and theoretical problems related to
T-odd distribution functions. 

\section{Origins of transverse spin asymmetries}

We will discuss possible origins of transverse spin asymmetries in the context
of the following hard scattering processes, that either have been or will 
be performed. 
First we will go into the details of the single and double
polarized Drell-Yan process  
$H_1 \,  H_2 \rightarrow \ell \, \bar\ell \, X$, for which there is no 
data available yet. 
Then we focus on the single polarized process $p \, p^{\uparrow} 
\rightarrow \pi \, X$ for
which large single transverse spin asymmetries have been observed
\cite{Adams}. We will also make use of knowledge of the 
{\em unpolarized\/} processes:
$\pi^- \, N \rightarrow \mu^+ \, \mu^- \, X$ and $e^+  \, e^- \rightarrow h_1
\,  h_2 \,  X$.

\subsection{The polarized Drell-Yan process}

Transverse spin asymmetries in hadron-hadron collisions require an
explanation that involves quarks and gluons. A large scale (the 
center of mass energy or the large lepton pair mass) allows for a 
factorization of such a process into parts describing the
soft physics convoluted with an elementary cross section. The parts 
parametrizing the soft physics cannot be calculated within perturbative QCD. 
Let us first focus on the
Drell-Yan process, i.e.\ lepton pair production in hadron-hadron collisions. 

\begin{figure}[htb]
\begin{center}
\leavevmode \epsfxsize=5cm \epsfbox{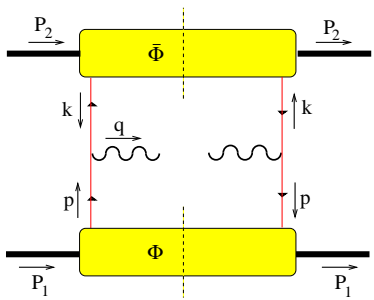}\\
\vspace{0.2 cm}
\caption{\label{LODY}The leading order contribution to the
Drell-Yan process}
\vspace{-5 mm}
\end{center}
\end{figure}

In lowest order, i.e.\ the parton model approximation, the Drell-Yan process 
consists of two soft parts (in Fig.\ \ref{LODY}
the leading order diagram is depicted \cite{Ralst-S-79}) 
and one of the soft parts is described
by the quark correlation function $\Phi(P_1,S_1;p)$ 
and the other soft part by the
antiquark correlation function, denoted by $\overline \Phi(P_2, S_2;k)$:
\ba
\Phi_{ij}(P_1,S_1;p)&=&\dz e^{ip \cdot z} 
\amp{P_1,S_1|\psibar_j (0) \psi_i (z) | P_1,S_1},\\
\overline \Phi_{ij}(P_2,S_2;k) & = & 
\int\frac{d^4z}{(2\pi)^4} \, e^{-i k\cdot z} 
\langle P_2,S_2 \vert \psi_i(z) \overline \psi_j(0) \vert P_2,S_2 \rangle.
\ea 
As will be discussed in the next section one can decompose the quark momenta
$p$ and $k$ into parts that are along the direction of the parent hadron,
the so-called lightcone momentum fractions, 
and deviations from that direction. In case one integrates over the transverse
momentum of the lepton pair 
one only has to consider the correlation 
functions as functions of the lightcone momentum fractions.   

The most general parametrization of the correlation function $\Phi$ as a
function of the lightcone momentum fraction $x$, that is in accordance with 
the required symmetries (hermiticity, parity, time reversal), is given by:
\beq
\Phi(x)=\frac{1}{2} \left[{f_1(x)} \mbox{$\not\! P\,$}+ 
{g_1(x)}\,\lambda\gamma_{5}\mbox{$\not\! P\,$}
+ {h_1(x)}\,\gamma_{5}\mbox{$\not\! S$}_{T}\mbox{$\not\! P\,$}\right].
\eeq
Other common notation is $q$ for $f_1$,  $\Delta q$ for $g_1$ and $\delta q $
or $\Delta_T q$ for $h_1$.

At this parton level one finds the well-known double transverse spin 
asymmetry \cite{Ralst-S-79},
\beq
A_{TT} \propto |\bm S_{1T}^{}|\;|\bm 
S_{2T}^{}|\;\cos(\phi_{S_1}+\phi_{S_2})\;
            h_1(x_1) \, \overline h_1(x_2),
\label{h1h1}
\eeq
which has not yet been experimentally observed, but is one of the objectives 
of the polarized
proton-proton scattering program to be performed at RHIC. This asymmetry
is one of the possible ways to get information on the transversity 
distribution function $h_1$.

At the parton model level there are no single transverse spin asymmetries,
but these might arise from corrections to this lowest order diagram. 
The corrections are of two types: the perturbative and the higher twist 
corrections. The first type depends logarithmically on the hard scale and the
second type behaves as inverse powers of the hard scale. 

Assuming that single spin asymmetries arise due to perturbative contributions
is conceptually the simplest option, since it assumes that the asymmetries
actually occur at the quark-gluon level, i.e.\ they arise from elementary
subprocesses, and that going to the hadron level
just involves convoluting the elementary asymmetry with (polarized) parton
distributions. Typically this will yield {\em single\/} transverse spin 
asymmetries 
of the order $\alpha_s m_q/\sqrt{s}$ \cite{Kane} which is expected to be 
small\footnote{Heavy
quarks will appear at higher
orders in $\alpha_s$ and have been shown to give rise to only small
contributions (even) to the unpolarized Drell-Yan cross section
\cite{Rijken}.}. The perturbative corrections to the {\em double\/} 
transverse spin asymmetry Eq.\ (\ref{h1h1}) 
have been calculated in \cite{Vogelsang-Weber} and using the
assumption that at low energies the transversity distribution function $h_1$
equals the helicity distribution function $g_1$, it has been shown in Ref.\
\cite{Martin} that $A_{TT}$ is expected to be of the order of a percent
at RHIC energies. 
We will view this as an indication that perturbative QCD contributions are
most likely not the (main) origin of large transverse spin
asymmetries. 
    
Dynamical higher twist corrections to the parton model 
require expanding the correlation function $\Phi(x)$ to
include contributions proportional to the hadronic scale (typically the hadron
mass), since these will show up in the cross section suppressed by $1/Q$,
where $Q$ is a hard scale. 
At leading order in $\alpha_s$, i.e.\ $(\alpha_s)^0$, but at the order
$1/Q$, one 
finds \cite{Jaffe-Ji-91,Tangerman-Mulders-95a} no single or double {\em
transverse\/} spin asymmetries\footnote{There 
is however a double spin asymmetry $A_{LT}$, which involves one
longitudinally and one transversely polarized hadron \cite{Jaffe-Ji-91}. A 
recent estimate of $A_{LT}$ using the bag model indicates that it is an
order of magnitude smaller than the leading order asymmetry $A_{TT}$ 
\cite{Kanazawa}.}. 

Hence, in order to produce a large single transverse spin asymmetry one
needs some conceptually nontrivial mechanism, since regular perturbative and
higher twist contributions appear to be either small or absent. 
Two such nontrivial mechanisms are the soft
gluon/fermion poles suggested by Qiu and Sterman \cite{QS-91b} and so-called
time-reversal (T) odd distribution functions (cf.\ e.g.\ \cite{Boer4}). 
Both of these mechanisms could produce
a single transverse spin asymmetry $A_T$ at order $1/Q$. 
Recently it has been shown \cite{Boer4}
that their effects are identical in the Drell-Yan process, so in order to
discriminate between them one must use other experiments as well. This 
asymmetry $A_T$ has been estimated to be of the order of a percent
at HERA energies (820 GeV, fixed target) \cite{Hammon-97}. 
Let us
remark that T-odd functions need not signal actual time reversal symmetry
violation. We
will discuss other options more extensively in the discussion at the end.  

In order to arrive at a single transverse spin asymmetry that is not 
suppressed by inverse powers of the hard scale, one can consider cross
sections differential in the transverse momentum of the lepton pair. In
that case one is sensitive to the transverse momentum of the quarks directly
and in case this concerns intrinsic transverse momentum of the quarks inside a
hadron, the effects need not be suppressed by $1/Q$. The point is that if the
transverse momentum of the lepton pair is produced by perturbative QCD
corrections, each factor of transverse momentum has to be accompanied by the
inverse scale in the elementary hard scattering subprocess, that is by
$1/Q$. But in case of {\em intrinsic\/} 
transverse momentum the relevant scale is not
$Q$, but the hadronic scale, say the mass of the hadron. In processes with two
(or more) soft parts, like the Drell-Yan process, the intrinsic transverse
momentum of one soft part is linked to that of the other soft part resulting
in effects, e.g.\ azimuthal asymmetries, not suppressed by $1/Q$. These
effects will show up at relatively low (including nonperturbative) 
values of $Q_T$, 
where $Q_T^2=\bm{q}_T^2$ and
$\bm{q}_T$ is the transverse momentum of the lepton pair. 
Studying the dependence of asymmetries on 
transverse momentum is another way to try to discriminate between possible 
origins for asymmetries. 

Returning to the parton model diagram and including intrinsic transverse
momentum dependence in this picture, one observes the following points. The
effects will only show up if $Q_T$ is observed (i.e.\ not integrated over).
If
only T-even structures are included, several double spin azimuthal asymmetries
are obtained, but no single spin asymmetries \cite{Tangerman-Mulders-95a}. 
So again one
needs to include some nontrivial mechanism. Gluonic and fermionic poles have 
as yet not been considered with transverse momentum dependence (other than
perturbatively produced), but would 
in any case appear in the cross section suppressed by a factor of $1/Q$.
However, the leading twist T-odd distribution 
functions with intrinsic transverse momentum 
dependence {\em do\/} yield single spin azimuthal asymmetries. 
We will be mainly focusing on the effects of such functions from now on.

\subsection{Pion production in $p \, p^{\uparrow}$ scattering}

\begin{figure}[htb]
\begin{center}
\leavevmode \epsfxsize=6cm \epsfbox{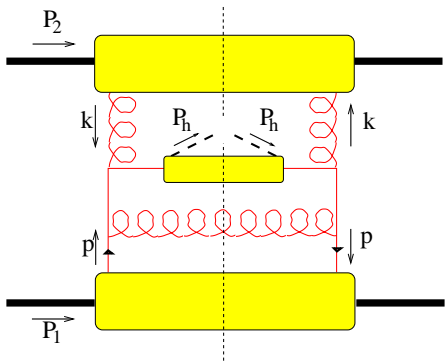}\\
\vspace{0.2 cm}
\caption{\label{pppiX}A contribution to the
process $p \, p^{\uparrow} \rightarrow \pi \, X$}
\vspace{-5 mm}
\end{center}
\end{figure}

The large single transverse spin asymmetries that have been observed 
in the process $p \, p^{\uparrow} \rightarrow \pi \, X$ \cite{Adams} require 
as said an
explanation that involves quarks and gluons. Again one needs large scales 
(in this case also a large transverse momentum of the pion) to allow for a 
factorization of this process into parts describing the
soft physics convoluted with an elementary cross section. For example, one
contribution is coming from the diagram depicted in Fig.\ \ref{pppiX}.

Assuming (as
argued above) that perturbative and higher twist corrections (gluonic and
fermionic pole contributions to this process have recently been investigated
in Ref.\ \cite{QS-98}) are too small to
generate the observed, large single transverse spin asymmetries, we will
restrict ourselves to the transverse momentum dependent T-odd functions, in 
this case both distribution and fragmentation functions. The so-called Sivers
\cite{s90} and Collins \cite{Collins-93b} effects are examples of transverse 
momentum dependent T-odd distribution and fragmentation functions, 
respectively. Like the transverse
momentum of the lepton pair in the Drell-Yan process, the transverse
momentum of the pion now originates from the intrinsic transverse momentum of
the initial partons in addition to transverse momentum perturbatively 
generated by radiating off some additional parton(s) in the final state. 

Anselmino {\em et al.\/} \cite{Anselmino} have investigated both the Sivers 
and the Collins
effect as possible origins for the asymmetries as observed in Ref.\
\cite{Adams}.  
Both effects can be used to fit the data, which then can be tested using other
observables. However, there are indications \cite{EST-98} 
from analyzing a particular angular dependence (a $\cos 2\phi$
dependence \cite{Boer}) in the unpolarized process 
$e^+ \, e^- \rightarrow Z^0
\rightarrow \pi \, \pi \, X$, where the pions belong to opposite jets,
that the Collins effect, is in fact at most a few percent of the magnitude of 
the ordinary unpolarized fragmentation function. 
Therefore, it seems unlikely that the Collins effect is the main source of the
single spin asymmetries of the $p \, p^{\uparrow} \rightarrow \pi \, X$
process.   

One other possible T-odd function that could be the source of the single spin
asymmetries is the chiral-odd function $h_1^\perp$, 
the distribution function analog of the Collins effect. It 
will be discussed extensively below for the case of the
Drell-Yan process, but it can equally well be the source of single spin
asymmetries in $p \, p^{\uparrow} \rightarrow \pi \, X$. On the other hand, 
the Collins effect itself will not contribute to the Drell-Yan process. 

\subsection{The unpolarized Drell-Yan process}  

The unpolarized cross section as measured for the 
process $\pi^- \, N \rightarrow \mu^+ \, \mu^- \, X$, where $N$ is either 
deuterium or tungsten, using a $\pi^-$-beam with energy of 
140, 194, 286 GeV \cite{NA10} and 252 GeV \cite{Conway},
\beq
\frac{1}{\sigma}\frac{d\sigma}{d\Omega} = \frac{3}{4\pi}\, 
\frac{1}{\lambda+3} \, \left( 1+ \lambda \cos^2\theta + \mu \sin^2\theta 
\cos\phi + \frac{\nu}{2} \sin^2 \theta \cos 2\phi \right),
\label{unpolxs}
\eeq
shows remarkably large values of $\nu$. It has been shown
\cite{NA10,bran93} that its magnitude cannot be explained by leading and 
next-to-leading order perturbative QCD corrections. A
number of explanations have been put forward, like a higher twist effect
\cite{bran94,Eskola}, which is the $1/Q^2$ 
term discussed by Berger and Brodsky \cite{Berger-80}. 
In Ref.\ \cite{bran94}
the higher twist effect is modeled using a pion distribution amplitude and it
seems to fall short in explaining the large values as found for $\nu$.
This higher twist effect would not be related to single spin asymmetries.

In Ref.\ \cite{bran93}
factorization breaking correlations between the incoming quarks are assumed
and modeled in order to account for the large $\cos 2\phi$ dependence. 
We will return to that extensively in section V.   
Another approach is put forward in Ref.\ \cite{Blazek} using 
coherent states. This can describe the $\cos 2\phi$ data, however, it fails 
to describe the function $\mu$ in a satisfactory manner.

From the point of view of transverse momentum dependent distribution functions
such a large $\cos 2\phi$ azimuthal dependence can arise at {\em leading
order\/} only from a 
product of two T-odd functions, in particular, only from the distribution 
function $h_1^\perp$.

We would like to mention the experimental observation that 
the $\cos 2 \phi$ dependence as observed by the NA10 collaboration
does not seem to show a strong dependence on $A$, i.e.\ there was
no significant difference between the deuterium and tungsten 
targets. Hence, it is unlikely that
it is in fact dominated by nuclear effects instead of effects associated
purely to hadrons. Therefore, the unpolarized cross section as can be measured 
at RHIC is also likely to show a large $\cos 2 \phi$ dependence, although
replacing the pion by a proton will probably have a suppressing effect. 

Hence, we conclude that although there exist, apart from the $h_1^\perp$ 
mechanism, several explanations of single
spin asymmetries and also of the unpolarized $\cos 2 \phi$ dependence in the
Drell-Yan cross section, none of the approaches relate the two types of
asymmetry and most of the
effects are expected or found 
to be (too) small. Moreover, the effects should not only
be large, they should also exhibit the right $Q_T$ behavior. 

\section{Transverse momentum dependent distribution functions}

In this section we will discuss the transverse 
momentum dependent distribution functions that are 
needed to find the expressions for
the leading order unpolarized and polarized 
Drell-Yan process cross sections differential in the transverse momentum of
the lepton pair. 

We consider again Fig.\ \ref{LODY}. The 
momenta of the quarks, which annihilate into the photon with momentum $q$, 
are predominantly along the direction of the parent
hadrons. One hadron momentum ($P_1$) is chosen to be along the 
lightlike 
direction given by the vector $n_+$ (apart from mass corrections). 
The second hadron with momentum $P_2$ 
is predominantly in the $n_-$ direction which satisfies $n_+ \cdot n_- = 1$, 
such that $P_1 \cdot P_2 = {\cal O} (q^2)$. 
We make the following Sudakov decompositions:
\begin{eqnarray}   
P_1^\mu &\equiv & \frac{ Q}{x_1 \sqrt{2}}\,n_+^\mu 
+ \frac{x_1 M_1^2}{ Q\sqrt{2}}\,n_-^\mu,\\
P_2^\mu &\equiv & \frac{x_2 M_2^2}{ Q\sqrt{2}}\,n_+^\mu
+ \frac{ Q}{x_2\sqrt{2}}\,n_-^\mu,\\
q^\mu &\equiv &\frac{ Q}{\sqrt{2}}\,n_+^\mu 
+ \frac{ Q}{\sqrt{2}}\,n_-^\mu 
+ q_T^\mu,
\end{eqnarray}
for $Q_T^2 \equiv - q_T^2 \equiv \bm{q}_T^2 \ll Q^2$. 
We will often refer to the $\pm$ components of a momentum $p$, 
which are defined 
as $p^\pm=p\cdot n_\mp$.
Furthermore, we decompose the parton momenta $p, k$ and the spin vectors 
$S_1, S_2$ of the two hadrons as
\ba
p & \equiv & \frac{x Q}{x_1 \sqrt{2}}\,n_+
+ \frac{x_1 (p^2 + \bm{p}_T^2)}{x Q \sqrt{2}}\,n_- + p_T
, \\[2 mm]
k & \equiv & \frac{\bar x Q}{x_2 \sqrt{2}}\,n_-
+ \frac{x_2 (k^2 + \bm{k}_T^2)}{\bar x Q \sqrt{2}}\,n_+ + k_T
, 
\\[2 mm]
S_1 & \equiv & \frac{\lambda_1 Q}{x_1 M_1 \sqrt{2}}\,n_+
-\frac{x_1 \lambda_1 M_1}{Q \sqrt{2}}\,n_- + S_{1T}
,\\[2 mm]
S_2 & \equiv & \frac{\lambda_2 Q}{x_{2} M_2 \sqrt{2}}\,n_-
-\frac{x_{2} \lambda_2 M_2}{Q \sqrt{2}}\,n_+ + S_{2T}^{}
.
\ea

The four-momentum conservation delta-function at the photon vertex is written 
as (neglecting $1/Q^2$ contributions)
\begin{equation} 
\delta^4(q-k-p)=\delta(q^+-p^+)\, \delta(q^--k^-)\, \delta^2(
\bm{p}_T^{}+
\bm{k}_T^{}-\bm{q}_T^{}),
\label{deltafn0}
\end{equation}
fixing 
$x P_1^+=p^+=q^+=x_1 P_1^+$, i.e., $x=x_1$ and similarly $\bar x = x_2$, and 
allows up to 
$1/Q^2$ corrections for integration over $p^-$ and $k^+$. 
However, the transverse momentum integrations cannot be separated, unless one 
integrates
over the transverse momentum of the photon or equivalently of the lepton pair.

The parametrization of $\Phi(p)$ should be 
consistent with requirements imposed 
on $\Phi$
following from hermiticity, parity and time reversal invariance. The latter is
normally taken to impose the following constraint on the correlation function 
\cite{Collins-93b,Tangerman-Mulders-95a} 
\begin{eqnarray}
&& \Phi^\ast(P,S;p) = \gamma_5 C \,\Phi(\bar P,\bar S;\bar p)\, 
C^\dagger \gamma_5 
\label{Tinvariance}
\end{eqnarray}
where $\bar p$ = $(p^0,-\bm{p})$, etc. 
For the validity of Eq.\
(\ref{Tinvariance}) it is essential that the incoming hadron is a plane wave
state. We will {\em not\/} 
apply this constraint and in the last section we will
discuss this issue in detail. 

In the calculation in leading order we encounter the correlation function 
integrated over $p^-$, which is parametrized in terms of the 
transverse momentum dependent distribution functions as \cite{Boer7}
\begin{eqnarray}
\Phi(x_1,\bm{p}_T) &\equiv & \left. \int dp^-\ \Phi(P_1,S_1;p) 
\right|_{p^+ = x_1 P_1^+, \bpt}
=
\frac{M_1}{2P_1^+}\,\Biggl\{
f_1(x_1 ,\bpt)\, \frac{\slsh{\! P_1}}{M_1} + 
f_{1T}^\perp(x_1 ,\bpt)\, \epsilon_{\mu \nu \rho \sigma}\gamma^\mu 
\frac{P_1^\nu p_T^\rho S_{1T}^\sigma}{M_1^2}
\nonumber \\[2 mm] 
\lefteqn{ \hspace{-13 mm} 
- g_{1s}(x_1 ,\bpt)\, \frac{\slsh{\! P_1} \gamma_5}{M_1}
- h_{1T}(x_1 ,\bpt)\,\frac{i\sigma_{\mu\nu}\gamma_5 S_{1T}^\mu P_1^\nu}{M_1}
- h_{1s}^\perp(x_1 ,\bpt)\,\frac{i\sigma_{\mu\nu}\gamma_5 p_T^\mu P_1^\nu}{M_1^2}
+ h_1^\perp (x_1,\bpt) \, \frac{\sigma_{\mu\nu} p_T^\mu
P_1^\nu}{M_1^2}\Biggl\}.}
\label{paramPhixkt}
\end{eqnarray}
We used the shorthand notation
\beq
g_{1s}(x_1, \bpt) \equiv
\lambda_1\,g_{1L}(x_1 ,\bm p_T^2)
+ \frac{(\bpt\cdot\bm{S}_{1T})}{M_1}\,g_{1T}(x_1 ,\bm p_T^2) 
\eeq
and similarly for $h_{1s}^\perp$.
The parametrization contains two T-odd functions, that would vanish if the
constraint Eq.\ (\ref{Tinvariance}) would be applied, i.e.\ the
Sivers effect function $f_{1T}^\perp$ and the analog of the Collins effect,
$h_1^\perp$. 

The parametrization of $\overline \Phi$ is
\begin{eqnarray}
\overline \Phi(x_2,\bm{k}_T) &\equiv & \left. \int dk^+\ \overline 
\Phi(P_2,S_2;k) 
\right|_{k^- = x_2 P_2^-, \bkt} = 
\frac{M_2}{2P_2^-}\,\Biggl\{ \overline f_1(x_2 ,\bkt)\,\frac{\slsh{\!
P_2}}{M_2}  
+ \overline f_{1T}^\perp(x_2 ,\bkt)\, 
\epsilon_{\mu \nu \rho \sigma}\gamma^\mu \frac{P_2^\nu
k_T^\rho S_{2T}^\sigma}{M_2^2}
\nonumber \\[2 mm]
\lefteqn{\hspace{-13 mm} 
+ \overline g_{1s}(x_2 ,\bkt)\, \frac{\slsh{\!P_2} \gamma_5}{M_2} 
- \overline h_{1T}(x_2 ,\bkt)\,
\frac{i\sigma_{\mu\nu}\gamma_5 S_{2T}^\mu P_2^\nu}{M_2}
- \overline h_{1s}^\perp(x_2 ,\bkt)\, 
\frac{i\sigma_{\mu\nu}\gamma_5 k_T^\mu P_2^\nu}{M_2^2}
+\overline h_1^\perp (x_2,\bkt) \, \frac{\sigma_{\mu\nu} k_T^\mu
P_2^\nu}{M_2^2} \Biggl\}.}
\end{eqnarray}

The Sivers effect function $f_{1T}^\perp$
has the interpretation of the distribution of an unpolarized quark with 
nonzero 
transverse momentum inside a transversely
polarized nucleon, while the function $h_1^\perp$ 
is interpreted as the distribution of a transversely polarized quark  
with nonzero 
transverse momentum inside an unpolarized hadron. In both cases the 
polarization is orthogonal to the transverse momentum of the quark.

In terms of these functions we can schematically say that in order to  
fit the $p \, p^{\uparrow} \rightarrow \pi \, X$ data, Anselmino 
{\em et al.\/} \cite{Anselmino} consider the following options for the
product of three functions that are parametrizing the three soft parts: 
$f_{1T}^\perp(x_1,\bpt) \overline f_1(x_2) D_1 (z)$
and $h_1(x_1) \overline f_1(x_2) H_1^\perp(z,\bkt)$, where $\overline
f_1(x_2)$ (or $D_1(z)$) can be the gluon distribution (or fragmentation) 
function $g(x_2)$ (or $G(z)$) instead also. 
However, there is one remaining option (for pion production), that we are
advocating as a source of single spin asymmetries: 
$h_1^\perp(x_1,\bpt) \overline h_1(x_2) D_1(z)$. Because of the
appearance of two chiral-odd quantities this
contribution might be expected to be 
smaller than $f_{1T}^\perp(x_1,\bpt) \overline f_1(x_2) D_1 (z)$. But even
though $f_1 \geq h_1$, one cannot exclude that 
$h_1^\perp(x_1,\bpt)$ is larger than $ f_{1T}^\perp(x_1,\bpt)$. 

Note that the magnitude of the Collins effect fragmentation 
function $H_1^\perp(z,\bkt)$ need not be
related to the magnitude 
of $h_1^\perp(z,\bkt)$. In contrast to the distribution
function, the fragmentation function will
receive contributions due to the final state
interactions which are present between the produced hadron and the
other particles produced in the fragmenting of a quark. This is also the
reason a similar constraint like Eq.\ (\ref{Tinvariance}) does not apply to
fragmentation correlation functions. 
  
\section{Unpolarized and single spin dependent cross sections}
 
The Drell-Yan cross section 
is obtained by contracting the lepton tensor with the hadron tensor 
\beq
{\cal W}^{\mu\nu}=\frac{1}{3} \int dp^- dk^+ d^2\bm{p}_T^{} d^2 
\bm{k}_T^{}\, \delta^2(\bm{p}_T^{}+
\bm{k}_T^{}-\bm{q}_T^{})\, \left.
\text{Tr}\left( \Phi (p) \, 
V_1^\mu \, \overline \Phi  
(k) \, V_2^\nu \right) \right|_{p^+, \, k^-}
+ \left(\begin{array}{c} 
q\leftrightarrow -q \\ \mu \leftrightarrow \nu
\end{array} \right).
\eeq
The vertices $V_i^\mu$ can be either the photon vertex 
$V^\mu = e \gamma^\mu$ or the
$Z$-boson vertex $V^\mu = g_V \gamma^\mu + g_A \gamma_5 \gamma^\mu$.
The vector and axial-vector couplings to the $Z$ boson are given by:
\begin{eqnarray} 
g_V^j &=& T_3^j - 2 \, Q^j\,\sin^2 \theta_W,\\
g_A^j &=& T_3^j,
\end{eqnarray} 
where $Q^j$ denotes the charge and $T_3^j$ the weak isospin of 
particle $j$ (i.e., $T_3^j=+1/2$ for $j=u$ and $T_3^j=-1/2$ for
$j=e^-,d,s$). 
We find for the leading order unpolarized Drell-Yan cross section, taking into
account both photon and $Z$-boson contributions, 
\ba
\lefteqn{
\frac{d\sigma^{(0)}(h_1h_2\to \ell \bar\ell X)}{d\Omega dx_1 dx_2 d^2{\bm 
q_T^{}}}=
\frac{\alpha^2}{3Q^2}\;\sum_{a,\bar a} \;\Bigg\{ 
          K_1(y)\;{\cal F}\left[f_1\overline f_1\right]}
\nonumber\\ && 
       \qquad + \left[K_3(y)\cos(2\phi)+K_4(y)\sin(2\phi)\right]\;
             {\cal F}\left[\left(2\,\bm{\hat h}\!\cdot \!
\bm p_T^{}\,\,\bm{\hat h}\!\cdot \! \bm k_T^{}\,
                    -\,\bm p_T^{}\!\cdot \! \bm k_T^{}\,\right)
                    \frac{h_1^{\perp}\overline h_1^{\perp}}{M_1M_2}\right]
\Bigg\}
\label{LO-OOO}
\ea
and for the case where hadron one is polarized:
\begin{eqnarray} 
\lefteqn{
\frac{d\sigma^{(1)}(h_1h_2\to \ell \bar\ell X)}{d\Omega dx_1 dx_2 d^2{\bm q_T^{}}}=
\frac{\alpha^2}{3Q^2}\;\sum_{a,\bar a} \;\Bigg\{ \ldots } 
\nonumber\\ && 
\qquad - {\lambda_1}\; \left[K_3(y)\sin(2\phi)-K_4(y)\cos(2\phi)\right] \;
             {\cal F}\left[\left(2\,\bm{\hat h}\!\cdot \!
\bm p_T^{}\,\,\bm{\hat h}\!\cdot \!
\bm k_T^{}\,
                    -\,\bm p_T^{}\!\cdot \!
\bm k_T^{}\,\right)
                    \frac{h_{1L}^{\perp}\overline h_1^{\perp}}{M_1M_2}\right]
\nonumber\\ && 
       \qquad + |\bm S_{1T}^{}|\;K_1(y)\;
\sin(\phi-\phi_{S_1})\; 
             {\cal F}\left[\,\bm{\hat h}\!\cdot \!\bm p_T^{}\,
                    \frac{f_{1T}^{\perp}\overline f_1}{M_1}\right]
\nonumber\\ && 
       \qquad - |\bm S_{1T}^{}|\;\left[K_3(y)\sin(\phi+\phi_{S_1})-
K_4(y)\cos(\phi+\phi_{S_1})\right]\;
             {\cal F}\left[\,\bm{\hat h}\!\cdot \!\bm k_T^{}\,
                    \frac{h_1\overline h_1^{\perp}}{M_2}\right]
\nonumber\\ &&
        \qquad - |\bm S_{1T}^{}|\;\left[K_3(y)\sin(3\phi-\phi_{S_1}) 
-K_4(y)\cos(3\phi-\phi_{S_1})\right]\;  \nonumber \\
&& \hspace{5 cm} \times {\cal F}\left[\left(
              4\,\bm{\hat h}\!\cdot \!\bm k_T^{}\,(\!\,
\bm{\hat h}\!\cdot \!\bm p_T^{}\,\!)^2
              -2\,\bm{\hat h}\!\cdot \!\bm p_T^{}\,\,
\bm p_T^{}\!\cdot \!\bm k_T^{}\,
              -\,\bm{\hat h}\!\cdot \!\bm k_T^{}\, \,
\bm p_T^2\,                    \right)
                 \frac{h_{1T}^{\perp}\overline h_1^{\perp}}{2M_1{}^2M_2}\right]
\Bigg\},\label{reducedO}
\end{eqnarray}
where the ellipsis stand for the T-even--T-even structures
(which for the contribution of the virtual photon are absent, cf.\ Ref.\ 
\cite{Tangerman-Mulders-95a}).  
Let us list the various definitions appearing in these expressions. 
We have defined the following combinations of the couplings and $Z$-boson 
propagators:
\ba
K_1(y)&=& A(y)\left[ e_a^2+ 2 g_V^l e_a g_V^a \chi_1 + c_1^l c_1^a \chi_2
\right] 
- \frac{C(y)}{2} \left[ 2 g_A^l e_a g_A^a \chi_1 + c_3^l c_3^a \chi_2 
\right],\\
K_2(y)&=& A(y)\left[ 2 g_V^l e_a g_A^a \chi_1 + c_1^l c_3^a \chi_2
\right] - \frac{C(y)}{2} 
\left[ 2 g_A^l e_a g_V^a \chi_1 + c_3^l c_1^a \chi_2 \right],\\ 
K_3(y)&=& B(y)\left[ e_a^2+ 2 g_V^l e_a g_V^a \chi_1 + c_1^l c_2^a \chi_2
\right], \\ 
K_4(y)&=& B(y)\left[ 2 g_V^l e_a g_A^a \chi_3 \right],
\ea
which contain the combinations of the couplings
\begin{eqnarray} 
c_1^j &=&\left(g_V^j{}^2 + g_A^j{}^2 \right),
\\[2 mm]
c_2^j &=&\left(g_V^j{}^2 - g_A^j{}^2 \right),
\qquad\qquad j=\ell\;\;\mbox{or}\;\;a
\\[2 mm]
c_3^j &=&2 g_V^j g_A^j.
\end{eqnarray}
The $Z$-boson propagator factors are given by
\ba
\chi_1 &=& \frac{1}{\sin^2 (2 \theta_W)} \, \frac{Q^2
(Q^2-M_Z^2)}{(Q^2-M_Z^2)^2 + \Gamma_Z^2 M_Z^2},\\
\chi_2 &=& \frac{1}{\sin^2 (2 \theta_W)} \, \frac{Q^2}{Q^2-M_Z^2} \chi_1,\\
\chi_3 &=& \frac{-\Gamma_Z M_Z}{Q^2-M_Z^2} \chi_1.
\ea

\begin{figure}[htb]
\begin{center}
\leavevmode \epsfxsize=9cm \epsfbox{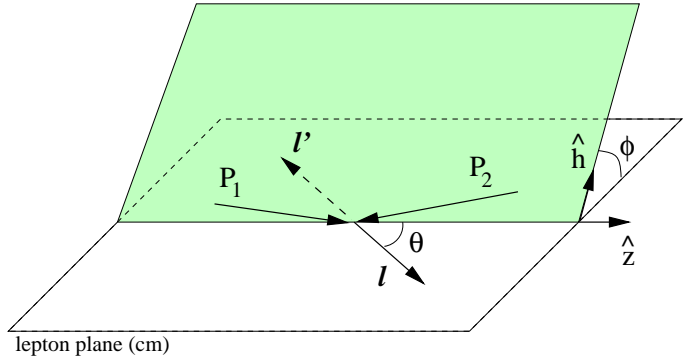}
\vspace{0.2 cm}
\caption{\label{DYkin} Kinematics of the Drell-Yan process in
the lepton center of mass frame.}
\vspace{-5 mm}
\end{center}
\end{figure}

The above is expressed in the so-called Collins-Soper frame \cite{CS}, 
for which
we chose the following sets of normalized vectors (for details see e.g.\
\cite{Boer4}):
\ba
\hat t &\equiv & q/Q,\\
\hat z &\equiv &\frac{x_1}{Q} 
\tilde{P_1}- \frac{x_2}{Q} \tilde{P_2},\\
\hat h &\equiv & q_T/Q_T = (q-x_1\, P_1 -x_2\, P_2)/Q_T,
\ea
where $\tilde{P_i} \equiv P_i-q/(2 x_i)$, such that:
\begin{eqnarray}
n_+^\mu & = & 
\frac{1}{\sqrt{2}} \left[ \hat t^\mu + \hat z^\mu
-\,\frac{Q_T^{}}{Q} \hat h^\mu \right], \label{nplusc}
\\
n_-^\mu & = & 
\frac{1}{\sqrt{2}} \left[ \hat t^\mu - \hat z^\mu
-\,\frac{Q_T^{}}{Q}\,\hat h^\mu \right].  \label{nplusc2}
\end{eqnarray}
The azimuthal 
angles lie inside the plane orthogonal to $t$ and $z$. In particular, 
$d\Omega$ = $2dy\,d\phi^l$, where $\phi^l$ gives the 
orientation of $\hat l_\perp^\mu \equiv \left( g^{\mu \nu}-\hat t^{ 
\mu} \hat t^{\nu } + \hat z^{
\mu} \hat z^{\nu } \right) l_\nu$, the perpendicular part of the lepton
momentum $l$; $\phi, \phi_{S_i}$ are the angles between $\bm{\hat h}, 
\bm S_{iT}^{}$ and $\hat l_\perp$, respectively. In the cross sections we also 
encounter the following functions of $y=l^-/q^-$, which in the
lepton center of mass frame equals $y=(1 + \cos \theta)/2$, where 
$\theta$ is the angle of $\hat z$ with respect to the momentum of the 
outgoing lepton $l$ (cf.\ Fig.\ \ref{DYkin}):
\ba
A(y) &=& \left(\frac{1}{2} -y+y^2\right) \stackrel{cm}{=} 
\frac{1}{4} \left( 1 + \cos^2\theta \right)
, \\
B(y) &=& y\,(1-y) \stackrel{cm}{=}\frac{1}{4} \sin^2 \theta 
,\\[2 mm]
C(y) &=& (1-2y) \stackrel{cm}{=} -\cos\theta.
\ea
Furthermore, we use the convolution notation 
(Ralston and Soper \cite{Ralst-S-79} use $I[...]$) 
\begin{equation} 
{\cal F}\left[f\overline f\, \right]\equiv \;
\int d^2\bm p_T^{}\; d^2\bm k_T^{}\;
\delta^2 (\bm p_T^{}+\bm k_T^{}-\bm 
q_T^{})  f^a(x_{1},\bm{p}_T^2) 
\overline f{}^a(x_{2},\bm{k}_T^2),
\end{equation}
where $a$ is the flavor index.
Since we are mainly concerned with the single  
polarized Drell-Yan process, we have given the double polarized cross section
in the appendix for completeness and future reference. 

\section{Quark spin correlations}

In order to explain the angular dependence of the 
unpolarized cross section as measured for the 
process $\pi^- \, N \rightarrow \mu^+ \, \mu^- \, X$, where $N$ is either 
deuterium or tungsten, using a $\pi^-$-beam with energy of 
140, 194 and 286 GeV \cite{NA10} ($-1/2 < \cos \theta < 1/2$), 
Brandenburg {\em et al.\/} \cite{bran93}
proposed factorization breaking correlations between the transverse momenta of
the incoming quarks and between their transverse spins. 
This correlation between the transverse momenta is taken to be
\beq
P(\bpt,\bkt) \, d^2\bpt \, d^2\bkt = \frac{\alpha_T (\alpha_T + 2\beta_T)}{ 
\pi^2} \, \exp \left[ -\alpha_T (\bm p_T^{2} + \bm k_T^2) - \beta_T (\bpt - \bkt)^2
\right]  \, d^2\bpt \, d^2\bkt,
\eeq
which reduces to separate Gaussian transverse momentum dependences, since 
$\beta_T$ is found to be practically zero (and 
$\alpha_T = 1 \, \text{GeV}^{-2}$ at these energies). 

In case the boson $V$ that produces the lepton pair, is a virtual photon 
($V=\gamma^*$) Brandenburg {\em et al.\/} 
fit the cross section Eq.\ (\ref{unpolxs}) 
using the data for the $\pi^-$-beam with energy of 194 GeV 
and lepton pair mass $m_{\gamma^*} = 8 \, \, \text{GeV}/c^2$.
They find that $1-\lambda -2\nu \approx -4 \kappa$, where 
they take the following 
model for $\kappa$, which is a measure of the correlation between the
transverse spins of the incoming quarks, 
\beq
\kappa =\kappa_0 \, \frac{Q_T^4}{Q_T^4 + m_T^4}
\label{kappaBMN}
\eeq
and the fitted values are $\kappa_0 =0.17$ and $m_T=1.5$ GeV. 

We will do a similar analysis based on the assumed 
presence of T-odd distribution functions with intrinsic transverse momentum
dependence. For simplicity we take 
$\mu=0$, $\lambda=1$ (in accordance with the expectation from next-to-leading
order perturbative QCD and 
the data in the Collins-Soper frame) and define $\nu = 2 \kappa$. 
For $V=\gamma^*$ we then find the following 
expression for $\kappa$ (cf.\ Eqs.\ (\ref{unpolxs}) and (\ref{LO-OOO})):
\beq
\kappa = \sum_{a,\bar a} e_a^2 \, 
{\cal F}\left[\left(2\,\bm{\hat h}\!\cdot \!
\bm p_T^{}\,\,\bm{\hat h}\!\cdot \! \bm k_T^{}\,
                    -\,\bm p_T^{}\!\cdot \! \bm k_T^{}\,\right)
                   \frac{h_1^{\perp}\overline h_1^{\perp}}{M_1M_2}\right]\Bigg/
\sum_{a,\bar a} e_a^2 \, {\cal F}\left[f_1\overline f_1\right].
\label{kappa1}
\eeq
A model for the shape of the function $h_1^\perp$ is needed. 
Collins' parametrization \cite{Collins-93b} for the fragmentation function 
$H_1^\perp$ is (note that Collins uses the function 
$\Delta \hat{D}_{H/a} \sim \epsilon_T^{ij} s_{1Ti} k_{T j} \, H_1^\perp$)
\beq
\frac{H_1^\perp(z, \bm{k}_T^2)}{D_1(z,\bm{k}_T^2)} = 
\frac{2 M_C M_h}{\bm{k}_T^2+ M_C^2} \, 
\text{Im}\left[A^*(k^2) B(k^2) \right] \,
\frac{(1-z)}{z},
\eeq
where $M_h$ is the mass of the produced hadron and in his model $M_C$ is the 
quark mass that appears in a dressed fermion propagator 
$i(A(k^2) \slsh{k} + B(k^2) M_C) /(k^2 -M_C^2)$, the functions $A$ and $B$ are
unity at $k^2=M_C^2$.
 
\begin{figure}[htb]
\begin{center}
\leavevmode \epsfxsize=9cm \epsfbox{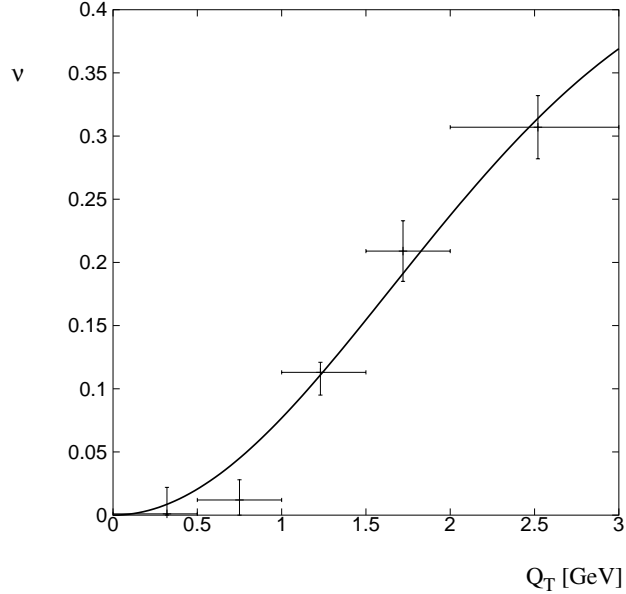}
\vspace{0.2 cm}
\caption{\label{fit}Data from \protect\cite{NA10} at 194 GeV and fit 
(using Eq.\ (\protect\ref{kappa2})) to $\nu=2\kappa$ as a function of the 
transverse momentum $Q_T$ of the lepton pair. The fitted parameters are 
$M_C=2.3 \pm 0.5 \, \text{GeV}$ and $16 \, \kappa_1 = 7 \pm 2$.}
\end{center}
\end{figure}

We assume a similar form for $h_1^\perp$ in terms of $f_1$ (we assume no
flavor dependence of $M_C$):
\beq
\frac{h_1^\perp{}^a(x, \bm{p}_T^2)}{f_1^a(x,\bm{p}_T^2)} = 
c_H^a \frac{M_C M_H}{\bm{p}_T^2+ M_C^2},
\eeq
using the constant $c_H^a$ (which in principle is a function of $x$) 
and $M_C$ as the fitting parameters now 
(and similarly for the antiquark distribution functions). We also assume the 
above given Gaussian transverse momentum dependence for $f_1(x,\bm{p}_T^2)$.
After multiplying Eq.\ (\ref{kappa1}) by a trivial factor $Q_T^2/Q_T^2$, 
using the $\bm{k}_T$ integration to eliminate the delta function and shifting
the integration variable
$\bm{p}_T \to \bm{p}_T'=\bm{p}_T^{} -{\scriptstyle {1\over 2}}\bm{q}_T^{}$, 
one arrives at
\beq
\kappa =  \frac{\kappa_1\, \alpha_T}{\pi}\, 
M_C^2 \, Q_T^2 \, \int d^2 \bm{p}_T' \,  
\left[ \frac{1}{(\bm{p}_T'+
{\scriptstyle {1\over 2}}\bm{q}_T)^2+ M_C^2}  
\, \frac{1}{(\bm{p}_T'-
{\scriptstyle {1\over 2}}\bm{q}_T)^2+ M_C^2}\right] 
\, e^{-2\alpha_T \bm p_T'{}^2},
\eeq
where $\kappa_1 =c_{H_1} c_{H_2}/2$ and for the moment we considered 
the one flavor case. 
We approximate this by taking $\bm{p}_T' =0$ (where the exponential factor is
largest) in the term between square brackets (this is valid for large enough
values of $Q_T$, but the resulting expression also has  
the right $Q_T^2$ behavior as $Q_T \to 0$), this results in
(reinstalling the flavor summation) 
\beq
\kappa =  8 \,  
\frac{Q_T^{2}\, M_C^2}{({Q}_T^2+ 4\,M_C^2)^2} \, 
\frac{\sum_{a,\bar a} e_a^2 \, \kappa_1^a \, f_1^a(x_1) \, 
\overline f{}_1^a(x_2) 
}{\sum_{a,\bar a} e_a^2 \, f_1^a(x_1) \, \overline f{}_1^a(x_2)}. 
\label{kappa3}
\eeq
Let us for simplicity also 
assume $\kappa_1^a$ to be independent of the flavor and fit 
\beq
\kappa =  8 \kappa_1 \,  
\frac{Q_T^{2}\, M_C^2}{({Q}_T^2+ 4\,M_C^2)^2} 
\label{kappa2}
\eeq
to the data at 194 GeV of Ref.\ \cite{NA10}. 
This does not give as good a fit (Fig.\ \ref{fit}) 
as a factor of $Q_T^4$ in the numerator would
give (for this particular set of data), 
but it can obviously reproduce the tendency. Moreover, it has the
desired property that $\kappa$ vanishes in the limit of $Q_T \to \infty$, as 
opposed to Eq.\ (\ref{kappaBMN}). We find for the dressed quark mass 
$M_C$ a rather large value of $2.3 \pm 0.5 \, \text{GeV}$ compared to the 
chiral symmetry breaking scale, but one should not take the model too 
seriously and we have made several approximations.   
We have chosen the data at 194 GeV of Ref.\ 
\cite{NA10}, because it has the smallest errors (the error in $Q_T$ is chosen
to be the bin size). 
The fits to the three other available sets of data, namely at 
140 and 286 GeV of Ref.\ 
\cite{NA10} and at 252 GeV of Ref.\ \cite{Conway}, yield  
lower values of $M_C$ and $\kappa_1$ (on average a factor 2 smaller), 
hence, have a lower maximum (at a
smaller value of $Q_T$) and are less
broad. We take the above result as providing a rough upper bound.

Taking for simplicity $c_{\pi}^a =c_{N}^a = c_{H}^a$, 
we arrive at a (crude) model for
the function $h_1^\perp(x_1, \bm{p}_T^2)$:
\beq
h_1^\perp{}^a(x, \bm{p}_T^2)= \frac{\alpha_T}{\pi}\, c_H^a\, 
\frac{M_C M_H}{\bm{p}_T^2+ M_C^2} \, e^{-\alpha_T \bm p_T^{2}} \, f_1(x),
\label{model}
\eeq
with $M_C=2.3 \, \text{GeV}$, $c_H^a=1$ and $\alpha_T =1\, \text{GeV}^{-2}$, 
which can be used to get rough estimates for other asymmetries. The factor
$\alpha_T/\pi$ 
comes from the consistency requirement between the definitions of $f_1(x)$ and
$f_1(x,\bm{k}_T^2)$ with a Gaussian $\bm{k}_T^2$ dependence. In the next
section we will discuss the relevant asymmetries for RHIC. 

\section{Implications for RHIC}

From Eq.\ (\ref{reducedO}) we see that in case $V=\gamma^*$ and that when 
we neglect the ``higher harmonic'' term containing the $3\phi$ dependence, 
there are two 
single transverse spin azimuthal dependences, namely $\sin(\phi-\phi_{S_1})$ 
arising with the Sivers function $f_{1T}^\perp$ 
and $\sin(\phi+\phi_{S_1})$ arising with $h_1^\perp$. 

To estimate the size of the $\sin(\phi-\phi_{S_1})$ term, one can
use for instance 
one of the usual parametrizations of $f_1$ and the parametrization for 
$f_{1T}^{\perp}$ as found by \cite{Anselmino}. To estimate the size of the 
$\sin(\phi+\phi_{S_1})$ term, one can use one of the models for $h_1$ 
\cite{Martin,Joao} or take the upper bound for $h_1$ that arises from 
Soffer's inequality ($g_1$ is also well-known) and one can use 
a fit for $h_1^\perp$ from the unpolarized azimuthal 
$\cos 2 \phi$ dependence of the cross section in $p\, p 
\rightarrow \ell \, \bar \ell \, X$, in a similar way as was 
done in the previous section.  

Let us examine the $\sin(\phi+\phi_{S_1})$ dependence 
of the cross section with the above given model  
for $h_1^\perp$. The relevant expression for the cross section in
the polarized case is given by (cf.\ Eq.\ (\ref{unpolxs}) with $\mu=0$ and 
$\lambda=1$)
\beq
\frac{1}{\sigma}\frac{d\sigma}{d\Omega\; d\phi_{S_1}} \propto 
\left( 1+ \cos^2\theta 
+ \kappa \; \sin^2 \theta \cos 2\phi - \rho \; |\bm S_{1T}^{}|\;
\sin^2 \theta\; \sin(\phi+\phi_{S_1}) + \ldots \right),
\label{polxs}
\eeq
where the ellipsis stand for the other angular dependences. 
The analyzing power $\rho$ is found to be (cf.\ Eq.\ (\ref{reducedO}))
\beq
\rho = \sum_{a,\bar a} e_a^2 \, 
{\cal F}\left[\bm{\hat h}\!\cdot \!
\bm k_T^{}\,\frac{h_1\overline h_1^{\perp}}{M_2}\right]\Bigg/
\sum_{a,\bar a} e_a^2 \, {\cal F}\left[f_1\overline f_1\right].
\label{rho1}
\eeq 
Using the above model Eq.\ (\ref{model}) for $h_1^\perp$ and performing 
similar approximations as before, we arrive at:
\beq
\rho = \frac{2 \, M_C \, Q_T}{Q_T^2+ 4 \, M_C^2} \, \frac{\sum_{a,\bar a} 
e_a^2 \, c_{H_2}^a \, h_1^a(x_1) \, \overline f{}_1^a(x_2) 
}{\sum_{a,\bar a} e_a^2 \, f_1^a(x_1) \, \overline f{}_1^a(x_2)}
=\frac{1}{2} 
\sqrt{\frac{\kappa }{\kappa_{\text{max}}}}\, \frac{\sum_{a,\bar a} 
e_a^2 \, c_{H_2}^a \, h_1^a(x_1) \, \overline f{}_1^a(x_2) 
}{\sum_{a,\bar a} e_a^2 \, f_1^a(x_1) \, \overline f{}_1^a(x_2)},
\eeq
where $\kappa_{\text{max}}$ is the maximum value of $\kappa$, which is at 
$Q_T=2 M_C$. 
A determination of $\rho, \kappa, h_1$ should be mutually consistent according
to the above equation,
if the underlying mechanism is indeed the one that is assumed here. The
maximum value $\rho$ is also at $Q_T=2 M_C$, which in the case of one flavor
corresponds to 
$\rho_{\text{max}} = c_H \, h_1(x_1)/(2 f_1(x_1)) \leq c_H/2 \approx 1/2$. 
If $h_1$ is for instance an order of
magnitude smaller than $f_1$, this would give an analyzing power for this
single transverse spin azimuthal asymmetry at the percent level. 

The above scheme entails many extrapolations and assumptions and prevents
us from stating accurate estimates for the asymmetries. One 
problem comes from the fact that the fit in the previous section resulted 
from data of the process $\pi^- \, N \rightarrow \mu^+ \, \mu^- \,  
X$, where $N$ is either deuterium or tungsten, so extrapolation to $p \, p 
\rightarrow \ell \, \bar \ell \, 
X$ is unclear. One might expect that the $\cos 2 \phi$
dependence of $p \, p 
\rightarrow \ell \,\bar \ell \, X$ as will be measured at RHIC is 
smaller than for the process $\pi^- \, N \rightarrow \mu^+ \, \mu^- 
\, X$, since in the former there are no valence antiquarks present.   
In this sense, the cleanest extraction 
of $h_1^\perp$ would be from $p \, \bar p \rightarrow \ell \, \bar \ell \, 
X$. 

Another problem concerns the energy scale. The extrapolations should involve 
evolving the functions to the 
relevant energies, however, the evolution equations for $f_{1T}^\perp$ and 
$h_1^\perp$ are not yet known.

However, the basic idea is clear. One fits the unpolarized azimuthal 
$\cos 2 \phi$ dependence of the cross section in a similar way as was  
done above, (for instance) by using a Collins' type of Ansatz to 
arrive at a model for $h_1^\perp$, which then can be used to measure or 
cross-check the 
function $h_1$ by measuring the $\sin(\phi+\phi_{S_1})$ dependence. 

\section{Discussion}

A $\cos 2 \phi$ term in the hadron tensor is itself a T-even quantity, but in
our approach it is factorized into a product of two T-odd functions. From the
definition of the correlation function $\Phi(p)$ 
one can show that time reversal
symmetry requires the T-odd functions to be zero \cite{Collins-93b}. 
This assumes that the incoming
hadrons can be describe as plane waves states. 
To circumvent this 
conclusion one could think of initial state interactions between the two
incoming hadrons \cite{Anselmino} or one could think of effects due to the 
finite size of a hadron \cite{Sivers-Zeuthen}. 

Initial state interactions 
between the two incoming hadrons would be a factorization breaking effect 
(not to be confused 
with the breakdown of factorization at higher twist \cite{Doria-80}) 
and this implies nonuniversality of the functions involved. 
The factorization breaking correlations
proposed by Brandenburg {\em et al.\/} \cite{bran93} assuming some 
nonperturbative gluonic background \cite{Ellis-Nachtmann}, 
might be universal in 
some restricted sense. For instance, one could retain universality among 
a subset of possible processes, namely the ones with exactly the same initial 
states. This would mean that 
functions obtained from the Drell-Yan process {\em can\/} be used to predict 
asymmetries in the process $p\, p \rightarrow \pi \, X$ successfully.
Another type of universality would be that the 
factorization breaking correlations are the same for different asymmetries
in the same process, e.g.\ the same for $\nu$ and $\rho$ in the case discussed
above. These issues can be tested experimentally. We
have proposed a concrete way to test some of these issues. 

At finite scales $Q_T$ and $Q$ one expects the finite size of a hadron to play
a role. However, such nonperturbative effects should not conflict with the 
factorization formula for Drell-Yan at finite $Q_T$ and 
$Q$ ($Q_T \ll Q$) \cite{coll85}. 
The finite size of hadrons most likely results in 
higher twist contributions, but maybe it will just prevent 
the naive application
of Eq.\ (\ref{Tinvariance}) as a constraint imposed by time reversal symmetry, 
which would not conflict with the factorization formula.  
These issues need to be investigated further theoretically. 

Let us just mention that finite size effects have been proposed 
as origins for the Sivers effect in Ref.\ \cite{Sivers-Zeuthen}. 
Spin-isospin interactions
have also been proposed \cite{adm96} to obtain a nonzero Sivers function. 
Liang {\em et al.\/} \cite{Liang-Boros} have proposed a model relating the
spin of a hadron to the orbital motion of quarks inside that hadron. This  
could be viewed as a model for the function $f_{1T}^\perp$ and a similar model 
might be constructed for the function $h_1^\perp$. 

It is worth emphasizing 
that the functions $f_{1T}^\perp$ and $h_1^\perp$ appear
in quite different asymmetries in general, even though they can both account
for the single spin asymmetries in $p \, p^{\uparrow} \rightarrow \pi \, X$. 
For instance, $f_{1T}^\perp$ cannot account for the $\cos 2 \phi$ asymmetry
discussed above and also, it yields a different angular dependence for the 
single spin asymmetry in the Drell-Yan cross section as was pointed out in the 
previous section. Also, in contrast to $h_1^\perp$, the Sivers effect, 
which is chiral-even, might produce 
single spin asymmetries in (almost) inclusive DIS 
\cite{Anselmino,adm96,Anselmino-Leader}, unless it originates from 
initial state interactions between hadrons.  

It is good to point out that the Berger-Brodsky higher twist mechanism is not
ruled out as a possible explanation for the observable $\nu$, although in the 
higher twist model of Ref.\ \cite{bran94} using a pion distribution amplitude 
it seems to fall short in explaining the large values found for $\nu$. Of
course, it might contribute
in addition to the $h_1^\perp$ mechanism. However, an observed 
correlation between $\nu$ and $\rho$ will be indicative of the latter. 

\section{Conclusions}
We have discussed in detail the consequences 
of T-odd distribution functions with intrinsic transverse momentum dependence 
for the Drell-Yan process. 
In particular, we focused on a chiral-odd T-odd distribution function,
denoted by $h_1^\perp$, which 
despite its conceptual problems, can in principle account for single spin
asymmetries in $p \, p^{\uparrow} \rightarrow \pi \, X$ and the Drell-Yan
process, and at the same time for the large 
$\cos 2\phi$ asymmetry in the unpolarized 
Drell-Yan cross section as found in Refs.\ \cite{NA10,Conway}, which still
lacks understanding.  
We have 
used the latter data to arrive at a crude model for this function and have 
shown explicitly how it relates unpolarized and polarized observables 
that could be studied at RHIC using polarized proton-proton collisions. 
It would also provide an alternative method of gaining 
information on the transversity distribution function $h_1$. 

The distribution function $h_1^\perp$ would signal a correlation 
between the transverse spin and the transverse momentum of quarks inside 
an unpolarized hadron. It is formally the distribution function analog of the
Collins effect, which concerns fragmentation, but most likely arises from
quite a different physical origin. Further theoretical and experimental 
study of these issues is required. 

We have also listed the complete set of azimuthal asymmetries in the 
unpolarized and polarized Drell-Yan 
process at leading order involving T-odd distribution functions with intrinsic
transverse momentum dependence.  

\acknowledgments 
I would especially like to thank Rainer Jakob and Piet Mulders for the past
collaborations on related topics, from which I greatly benefited. 
Also, I thank Alessandro Drago, Hirotsugu Fujii, Xiangdong Ji, Amarjit Soni, 
Oleg Teryaev and Larry Trueman for valuable discussions. 
The cross sections are obtained with FORM and the plot and fit with Gnuplot3.7.

\section*{Appendix}

The leading order double polarized Drell-Yan cross section, taking into
account both photon and $Z$-boson contributions, is found to be
\begin{eqnarray} 
\lefteqn{
\frac{d\sigma^{(2)}(h_1h_2\to \ell \bar\ell X)}
     {d\Omega dx_1 dx_2 d^2{\bm q_T^{}}}=
\frac{\alpha^2}{3Q^2}\;\sum_{a,\bar a} \;\Bigg\{ \ldots }
      \nonumber\\ &&
        + \frac{K_1(y)}{2}\;|\bm S_{1T}^{}|\;
                         |\bm S_{2T}^{}|\;
                \cos(2\phi-\phi_{S_1}-\phi_{S_2})\;
             {\cal F}\left[\,\bm{\hat h}\!\cdot \!\bm p_T^{}\,
                    \,\bm{\hat h}\!\cdot \!\bm k_T^{}\,
                    \frac{f_{1T}^\perp \overline f_{1T}^\perp
                         -g_{1T} \overline g_{1T}}{M_1M_2}\right]
\nonumber\\ &&
        - \frac{K_1(y)}{2}\;|\bm S_{1T}^{}|\;
                          |\bm S_{2T}^{}|\;
                \cos(\phi-\phi_{S_1})\;\cos(\phi-\phi_{S_2})\;
             {\cal F}\left[\,\bm p_T^{}\!\cdot \!
                      \bm k_T^{}\,
                    \frac{f_{1T}^\perp \overline f_{1T}^\perp}{M_1M_2}\right]
\nonumber\\ &&
        - \frac{K_1(y)}{2}\;|\bm S_{1T}^{}|\;
                          |\bm S_{2T}^{}|\;
                \sin(\phi-\phi_{S_1})\;\sin(\phi-\phi_{S_2})\;
             {\cal F}\left[\,\bm p_T^{}\!\cdot \!
                      \bm k_T^{}\,
                    \frac{g_{1T} \overline g_{1T}}{M_1M_2}\right]
\nonumber\\ &&
     +K_2(y)\;\lambda_1\;|\bm S_{2T}^{}|\;
                 \sin(\phi-\phi_{S_2})\;
          {\cal F}\left[\,\bm{\hat h}\!\cdot \!\bm k_T^{}\,
                 \frac{g_1 \overline f_{1T}^\perp}{M_2}\right]
\nonumber\\ && 
     +K_2(y)\;|\bm S_{1T}^{}|\;
                                 |\bm S_{2T}^{}|\;
          \sin(2\phi-\phi_{S_1}-\phi_{S_2})\;
          {\cal F}\left[\,\bm{\hat h}\!\cdot \!\bm p_T^{}\,
                 \,\bm{\hat h}\!\cdot \!\bm k_T^{}\,
                 \frac{f_{1T}^\perp \overline g_{1T}}{M_1M_2}\right]
\nonumber\\ &&
     -K_2(y)\;|\bm S_{1T}^{}|\;
                                 |\bm S_{2T}^{}|\;
          \cos(\phi-\phi_{S_1})\sin(\phi-\phi_{S_2})\;
          {\cal F}\left[\,\bm p_T^{}\!\cdot \!
                   \bm k_T^{}\,
                 \frac{f_{1T}^\perp \overline g_{1T}}{M_1M_2}\right]
      +\quad {\;1\longleftrightarrow 2\;\choose 
              \;\bm{p}\longleftrightarrow\bm{k}\;} \quad
\Bigg\},
\end{eqnarray} 
where the ellipsis stand for the (remaining) T-even--T-even structures
(which for the contribution of the virtual photon can be found in Ref.\ 
\cite{Tangerman-Mulders-95a}).

\end{document}